# Tunable Plasmonic Toroidal Terahertz Metamodulator

Burak Gerislioglu, Arash Ahmadivand, *Member, IEEE* and Nezih Pala, *Senior Member, IEEE*

*Abstract*—**Optical modulators are essential parts of photonic circuits to encode electro-optical signals to the optical domain. Here, using arrays of multipixel toroidal plasmonic terahertz (THz) metamolecules, we developed a plasmonic metamodulator with high efficiency and tunability. Using the ultrasensitivity of the excited toroidal momentum to the incident THz beam power, we numerically and experimentally studied the plasmonic response of the proposed device. We showed that the proposed THz toroidal metamodulator have strong potential to be employed for practical advanced communication applications.**

*Index Terms*—**Terahertz plasmonics, toroidal moments, metamaterial, modulation.**

## I. INTRODUCTION

TERAHERTZ (THz) plasmonic metamaterials consisting of periodic arrays of artificially tailored subwavelength blocks with hybrid nature and capability to sustain ultrastrong plasmonic resonant momenta [1,2], have been extensively employed for developing ultrafast and efficient metadevices including but not limited to active phase modulators [3-5], tunable filters [6,7], interferometers [8,9], biomarkers sensors [10-12], microelectromechanical systems (MEMS) [13], THz transistors and detectors [14,15]. The ability to support pronounced high quality plasmonic lineshapes (i.e. Fano resonances [16-19], electromagnetically induced transparency (EIT) [20,21], and charge transfer plasmons (CTPs) [22-24]) allowed for tailoring advanced photonic devices. On the other hand, recently, a third family of nonradiative resonant modes has been excited successfully using plasmonic metamolecules known as toroidal moments with weak far-field optical patterns due to masking by dominant classical electromagnetic multipolar far-field radiation [25-27]. Technically, possessing unconventional dynamic multipole expansion of current excitation, toroidal multipoles basically function based on both time-reversal ($t \rightarrow -t$) and space inversion ($r \rightarrow -r$) symmetry, which are the fundamentals of toroidization concept in nuclear physics [28,29]. The strongest member of toroidal multipoles is the hybrid toroidal dipole component, which can be identified as a vortex of closed-loop circular magnetic current flowing across the surface of a torus along the corresponding meridian [25,27,30]. In the toroidal spectral feature limit, the electromagnetic field can be substantially squeezed in a single spot due to the lack of coupling of toroidal radiation to free space. In particular, such an exquisite inherent feature of toroidal metamolecules allows for achieving exotic effects such as strong field localization [31], and cloaking [32]. Recent accomplishments in submillimeter and THz plasmonic metamaterials technologies have revealed the excitation of ultrastrong and narrow linewidth toroidal dipoles using both planar [26,27,33] and artificial three-dimensional (3D) meta-atoms [25,26,34]. Despite of rapid and impressive progress in THz toroidal metasurface technology, the induced resonant modes suffer from lack of tunability. Conversely, it is shown that Fano- and EIT-based THz metasurfaces show substantial spectral response tunability and have opened new doors for developing functional plasmonic devices [35-37]. For instance, the required tunability was achieved by integrating the systems with optoelectronically controllable materials (i.e. graphene) and also by thermal actuation. Although most of these techniques facilitate are promising for designing practical devices, they suffer from complex and expensive fabrication processes.

Comparing to the classical resonant modes, toroidal moments show more sensitivity to the morphological, incident beam power, and environmental refractive index variations [12]. Therefore, achieving tunable toroidal metamaterials allows for developing highly sensitive optical devices. In this Letter, using the strong dependency of spectral response of toroidal metamolecules to incident THz wave power, we designed, fabricated and extensively studied the spectral properties of a toroidal plasmonic metamodulator with high responsivity, significant figure of merit (FoM) and negligible insertion losses. To this end, a plasmonic resonator composed of gold blocks were employed to induce toroidal momentum with high quality factor ($Q$ factor). Using numerical predictions, following by experimental analysis, we demonstrated fast and functional THz plasmonic on/off modulator, controllable by incident beam power.

## II. EXPERIMENTAL AND NUMERICAL METHODS

For the fabrication of the proposed planar metamolecules, we

This manuscript is submitted by Dec 13, 2017. This work is supported by Army Research Laboratory (ARL) Multiscale Multidisciplinary Modelling of Electronic Materials (MSME) Collaborative Research Alliance (CRA) (Grant No. W911NF-12-2-0023, Program Manager: Dr. Meredith L. Reed).

A. Ahmadivand is with the Department of Electrical and Computer Engineering, Florida International University, 10555 W Flagler St, Miami, FL 33174, United States. (*Corresponding Author*, email: aahma011@fiu.edu).
B. Gerislioglu and N. Pala are with the Department of Electrical and Computer Engineering, Florida International University, 10555 W Flagler St, Miami, FL 33174, United States.

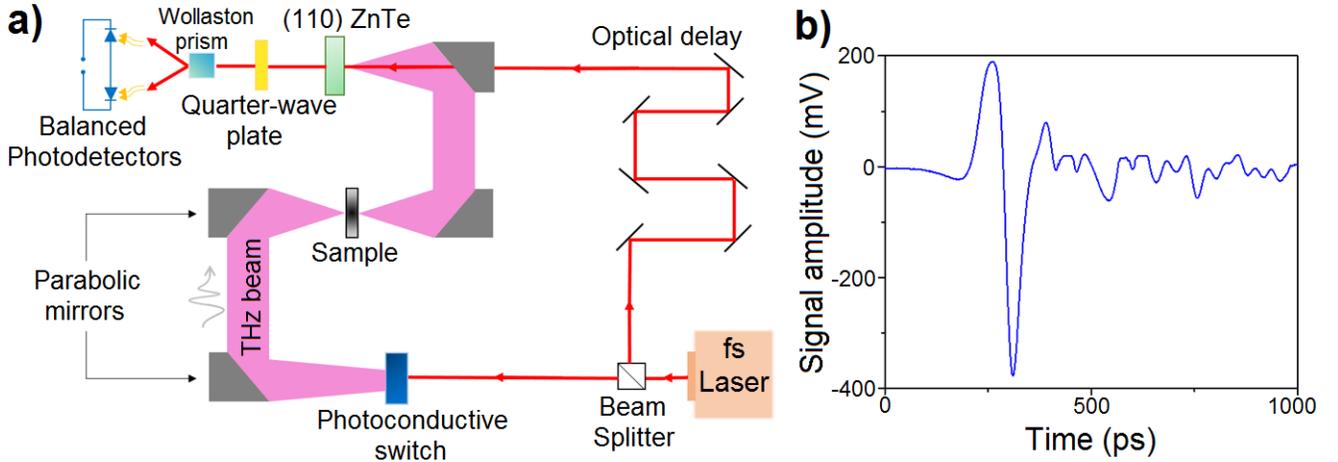

Figure. 1. a) Schematic representation of the THz-TDS setup. b) The applied electro-optical signal amplitude as a function of time delay.

carried out single-level lithography microfabrication process. An undoped and high-resistivity silicon (Si) wafer (>10 kΩ.cm) with the crystal orientation of <100> and thickness of 500 μm was used as substrate to provide the required transparency in the THz spectra [24]. After sonication, we deposited negative photoresists (NLOF 2020) with the thickness of 2 μm and patterned in multiple steps. Employing e-beam evaporation, we then deposited 50 nm of titanium (Ti) and 300 nm of gold (Au) layers separately with the rate of 1 Å/s (99.99% purity for Ti and 99.9995% purity for Au, vacuum pressure ~3.2×10$^{-4}$ mTorr). The Ti sublayer was used to enhance the adherence of gold layer to the surface of the Si wafer. The THz characterization was performed using a THz-TDS setup with the beam bandwidth of 10 GHz to 4.5 THz with varying power (Fig. 1a). Figure 1b illustrates the TDS waveform. For the numerical modelling, the finite-difference time-domain (FDTD) approach (Lumerical 2018) was used with the PML layers as the workplace boundaries for radiation direction (z-axis) and periodic boundaries for x and y-axes and a plane wave pulse served as a THz source. The dielectric function of simulated gold was taken from the empirically defined values by Ordal *et al.* [38].

### III. RESULTS AND DISCUSSION

Figure 2a illustrates an array of our proposed plasmonic metamolecules on a silicon substrate, showing the polarization of the incident THz beam. The corresponding geometrical parameters for the proposed toroidal unit cell are specified in Fig. 2b. We used following sizes for the structure geometries: $R_i/W/L/D/g$=60/15/105/15/5 μm. The scanning electron microscope (SEM) graph of the metamolecule arrays is depicted in Fig. 2c. For the incident y-polarized THz radiation, we calculated and plotted the transmission amplitude spectra for various beam powers as shown in Figs. 3a and 3b, using FDTD and TDS analyses, respectively. Obviously, for the beam with maximum power (12 mW), a distinct dipolar momentum correlating with the toroidal spectral feature is excited around ~0.18 THz. Then, we gradually reduced the beam power down to 0 mW. In this regime, the appeared toroidal momentum progressively decays and finally disappears, resulting in a transparent THz metamaterial. Such a dependency on the incident beam power can be exploited for developing a plasmonic toroidal modulator. Figures 3c to 3e help to understand the physical mechanism behind the formation of toroidal momentum and its behavior to the incident beam variations. The local near-field maps in Fig. 3c demonstrate dramatic decay in the intensity of localized plasmons, when the power is reducing. According to the toroidization concept, the localized plasmons in structures with gaps play fundamental role in determining the direction of surface current and formation of head-to-tail magnetic moments [27]. As shown in the E-field snapshots, reducing the power of the beam gives rise to drastic decay in capacitive gap plasmons. For instance, for the lowest applied pump power (0 mW), the gap plasmons entirely damped due to the decay in the surface currents in both resonators. Figure 3d exhibits cross-sectional plane for the induced magnetic surface current as a function of position in the $x$-axis at the toroidal dipole resonant mode, reaching up to ~$6.5 \times 10^5$ A/m. This profile implies that stronger surface currents leads to formation of intense mutual inductance between proximal resonators and results in significant confinement of plasmons in the metamolecules arrays. The inset schematically shows the generation of opposite surface currents and the induced toroidal dipolar moment in the proposed metamolecule based on the calculated the vectorial charge density presented in Fig. 3e. Hence, reducing the power of the incident beam dramatically affects the induced surface current across the unit cell, leading to active control of the intensity and quality of the toroidal spectral feature. In continue, we use this advantage of toroidal momentum to develop an ultrafast, and efficient plasmonic THz metamodulator with high responsivity.

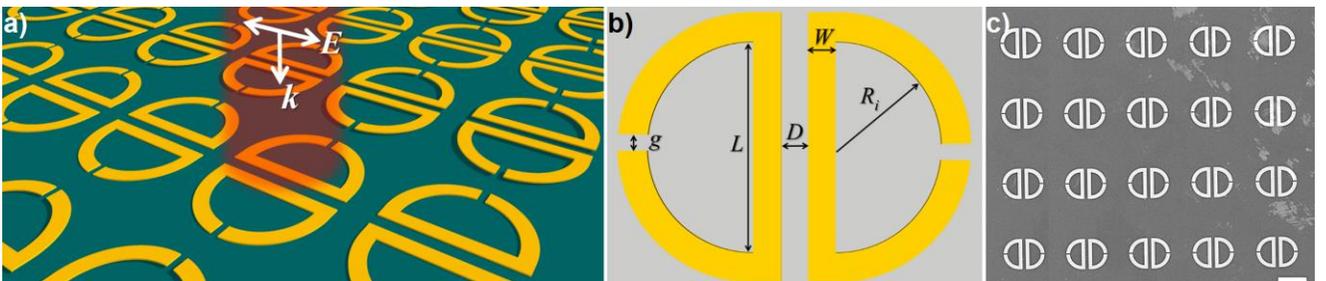

Figure. 2. a) and b) Schematic representation of the toroidal metamolecule arrays and the specified geometrical parameters, respectively. c) The SEM image of fabricated metamolecules array. The scale bar in this image is 100 μm.

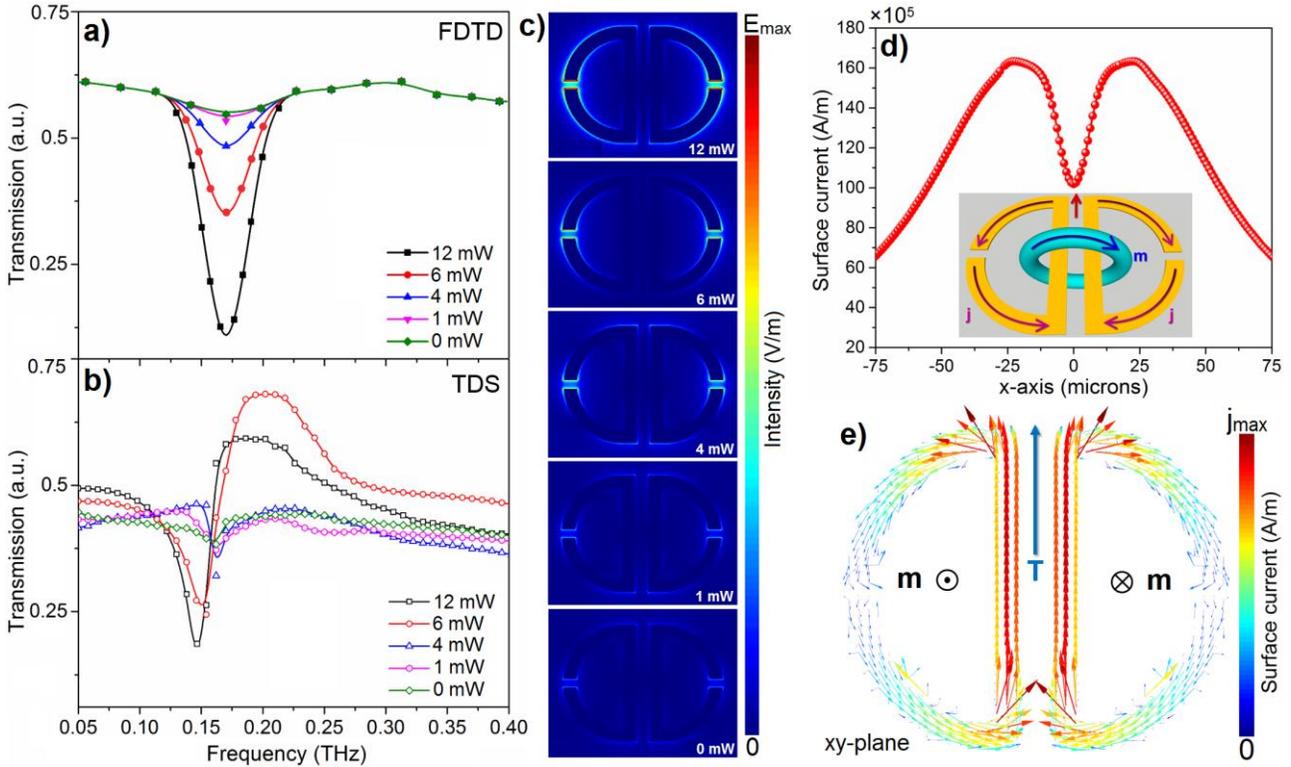

Figure. 3. a), and b) Numerically (FDTD) and experimentally (TDS) obtained transmission amplitude spectra for the metamolecule, respectively, while the power of the incident THz beam is varying. c) The local E-field maps for the unit cell for varying incident power. d) Cross-sectional surface current density as a function of position. e) A vectorial map for the current oscillation at the toroidal dipole frequency.

Figure 4 contains the experimental results for the switching properties of the proposed metamodulator. First, we analyzed the $Q$ factor of the induced toroidal dipolar momentum. Quantifying this parameter allows defining the other key parameters. The highest experimental (TDS) $Q$ factor (29.8) is obtained for the maximum applied pump power of 12 mW. Further reduction in the beam power reduced the $Q$ factor to almost zero for 0 mW of incident beam. Similar scenario is observed for the surface current across the unit cell, obtained by applying current module in FDTD program. The reduced surface current is a strong proof for the decay in the toroidal spectral feature. The inset diagram demonstrates the switching properties of the proposed plasmonic device. Here, we quantified the resultant insertion loss ($IL$ [dB]=-10Log₁₀ ($|T_{max}|$)) of the THz metarouter as a function of THz beam power. In this regime, increasing the beam power reduces the $IL$ factor and enhances the quality of the device. The lowest $IL$ ~4.25 dB was obtained for the maximum power. The inset diagram in Fig. 4 also includes the figure of merit (FoM), computed using the relation between the $IL$ and extinction ratio (-10Log₁₀ ($|T_{min}|/|T_{max}|$)), given by [39]:

$$\text{FoM} = \frac{Log_{10}\left(\left|\frac{T_{min}}{T_{max}}\right|\right)}{Log_{10}\left(\left|T_{max}\right|\right)} \tag{1}$$

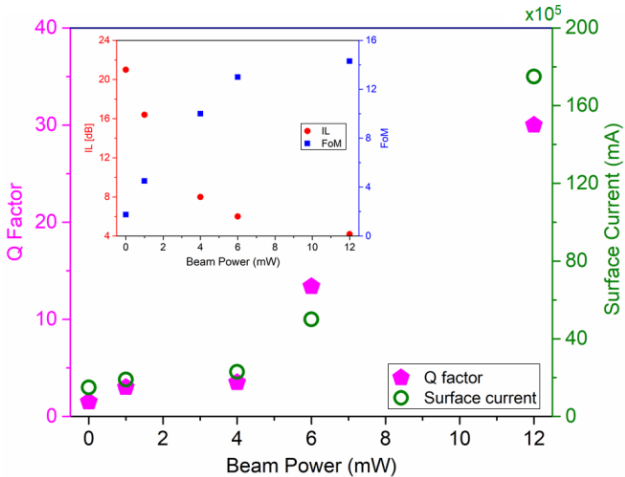

Figure. 4. The toroidal dipole Q factor and induced surface current as a function of beam power. The inset is the $IL$ and FoM of toroidal spectral feature as a function of beam power.

The FoM of the proposed metamodulator as a function of the incident beam power reflects increase of the FoM for the increasing beam power. The maximum FoM for the highest incident power is found as 15.4 for 12 mW.

## IV. CONCLUSIONS

Here, we have developed a plasmonic modulator for THz frequencies by taking advantage of the strong dependency of the toroidal dipolar mode on the incident beam power. Our numerical studies confirmed with the experimental results showed that an incident THz beam with a high enough power induces toroidal moments in the proposed metasurface resulting in a distinct resonance line shape in the transmission spectrum. The Quality of the resonance decreases with the reduced incident power and the peak disappears below the threshold power. The demonstrated device can be used in advanced THz photonics applications.